\documentclass[structabstract]{aa}  
\usepackage{graphicx}
\usepackage{txfonts}
\usepackage{longtable,lscape}

\def\variables{350 }
\def\varisa{173 }
\def\varisb{177 }

\begin{document}
   \title{Periodic variable stars in CoRoT field LRa02 observed with BEST II\footnote{Tables 4\&5 and figures 5\&6 are online only.}}

   \subtitle{}

   \author{P. Kabath\inst{1}
          \and
          A. Erikson\inst{1}
          \and
          H. Rauer\inst{1}\inst{2}
          \and
          T. Pasternacki\inst{1}
          \and
          Sz. Csizmadia\inst{1}
          \and
          R. Chini\inst{3}
          \and
          R. Lemke\inst{3}
          \and
          M. Murphy\inst{4}
          \and
          T. Fruth\inst{1}
          \and
          R. Titz\inst{1}
          \and
          P. Eigm\"{u}ller\inst{5}}

   \institute{Institut f\"ur Planetenforschung, Deutsches Zentrum f\"ur Luft- und Raumfahrt, Rutherfordstr. 2, 12489 Berlin, Germany\\
              \email{petr.kabath@dlr.de}
         \and
             Zentrum f\"ur Astronomie und Astrophysik, Technische Universit\"at Berlin, 10623 Berlin, Germany\\
         \and
             Astronomisches Institut, Ruhr-Universit\"{a}t Bochum, 44780 Bochum, Germany\\
         \and
             Depto. F\'{i}sica, Universidad Cat\'{o}lica del Norte
P.O. 1280, Antofagasta, Chile\\
         \and
             Th\"{u}ringer Landessternwarte Tautenburg, 07778 Tautenburg, Germany            
             }

   \date{Received 20 February 2009 /Accepted 24 June 2009  }

 
  \abstract
   {The Berlin Exoplanet Search Telescope II (BEST II) is a small wide field-of-view photometric survey telescope system located at the Observatorio Cerro Armazones, Chile. The high duty cycle combined with excellent observing conditions and millimagnitude photometric precision makes this instrument suitable for ground based support observations for the CoRoT space mission.}
   {Photometric data of the CoRoT LRa02 target field collected between November $2008$ and March $2009$ were analysed for stellar variability. The presented results will help in the future analysis of the CoRoT data, particularly in additional science programs related to variable stars.}
   {BEST II observes selected CoRoT target fields ahead of the space mission. The photometric data acquired are searched for stellar variability, periodic variable stars are identified with time series analysis of the obtained stellar light curves.}
   {We obtained the light curves of $104335$ stars in the CoRoT LRa02 field over $41$ nights. Variability was detected in light curves of $3726$ stars of which \variables showed a regular period. These stars are, with the exception of $5$ previously known variable stars, new discoveries.}
   {}

   \keywords{methods: data analysis --- binaries: eclipsing -- stars: variables: general}

   \maketitle
%

\section{Introduction}

The Berlin Exoplanet Search Telescopes (BEST and BEST II) are two small-aperture wide-field telescope systems operated as ground-based support facilities for the CoRoT space mission (\cite{baglin}). The first telescope, BEST (\cite{Rauer2004}, \cite{Rauer2009}), operational since $2004$, is located at the Observatoire de Haute Provence, France. The second telescope, BEST II, has been operational since April $2007$ at Observatorio Cerro Armazones, Chile. Both telescopes are used to perform preparatory ground-based variability characterization of the target fields of the space mission. Normally the target fields are observed one year ahead of CoRoT.  

The CoRoT mission will complete observations of five long-run fields (150 days) and several shorter runs during the nominal mission phase (\cite{Deleuil2006}). In support of the space mission, variability characterization of three long run and one initial run fields have been performed within the BEST project (\cite{Karoff2007}, \cite{Kabath2007}, \cite{Kabath2008}, \cite{Kabath2009}), light curves from CoRoT later can be extended with BEST observational data. Thus, significant contributions can be made to the CoRoT additional scientific programs related to variable stars. Usually, several hundred new variable stars of various types ($\beta$ Lyr, $\delta$ Sct, Algol type eclipsing binaries, $\delta$ Cep, RR Lyr and others) are detected easily with the BEST telescopes in the CoRoT fields. In addition, the data from the BEST telescopes can be used during the confirmation process of CoRoT planets, as exemplified by the BEST pre-discovery transit observations of CoRoT 1b and CoRoT 2b (\cite{Rauer2009}).

The fourth CoRoT long-run field LRa02 was observed with BEST II during the Chilean winter period $2007/2008$, one year prior to the space mission. The acquired light curves were analyzed for stellar variability. \cite{poretti} have reported new variable stars in the same field. However, their detected variable stars are brighter than the cases presented here. Thus, the BEST II observations of the LRa02 field are complementary to already published results and within the same magnitude range as the CoRoT observations. The detected stars presented here were cross matched with the 2MASS catalog and the light curves of a few known periodic variable stars in the field could be extended by our data.

\begin{figure}
   \centering
  \includegraphics[width=8cm, height=8cm]{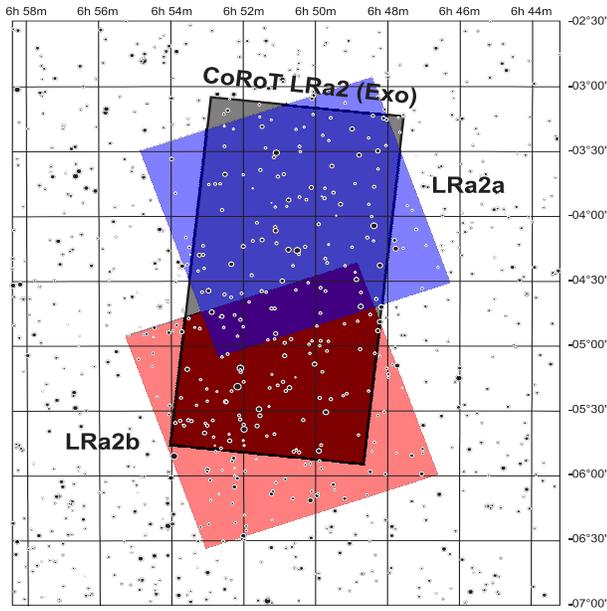}
      \caption{The orientation of BEST II LRa02 subfields with respect to CoRoT's LRa2b field (coordinates J2000.0).}
         \label{figfo}
   \end{figure}

\section{Telescope and observational campaign}

The BEST II telescope is located at the Observatorio Cerro Armazones, Chile, and operated in robotic mode. The system consists of a $f/5.0$ Baker-Ritchey-Chr\'{e}tien telescope with $25$-cm aperture and is equipped with a Peltier cooled $4$K$\times4$K, $16$ bit Finger Lakes Imager CCD of type KAF-16801E1 with a $9\mu$m pixel size. The peak quantum efficiency reaches $68\%$ at $650$nm. The field of view (FOV) of the system covers $1.7^\circ\times 1.7^\circ$ on the sky with an angular resolution of $1.5"pixel^{-1}$. BEST II observes without filter and with a light sensitivity of the CCD that is approximately equivalent to a broad $R$ band. The commercial mount, Atacama GM-4000, manufactured by 10 Micron, allows high precision pointing. To support the observations, a meteorological weather station and webcams monitor the conditions at the site. 

We observed the CoRoT LRa02 field from the end of November $2008$ to the end of February $2009$ for $41$ nights. The whole campaign lasted $91$ days. High precision pointing allowed us to alternate between two BEST II fields, thereby covering the whole CoRoT exoplanetary FOV (see Figure~\ref{figfo}). The center coordinates of the two observed BEST II subfields are:\\

\begin{center} 
LRa2a: $\alpha = 06^h 50^m 46.3^s,\; \delta=-03^\circ 59' 31.0"$\\  
LRa2b: $\alpha = 06^h 51^m 13.9^s,\; \delta=-05^\circ 26' 16.0"$. 
\end{center}

The observational sequence consists of $240$ sec exposures followed by dark and bias frames taken approximately each hour. In addition, standard calibration images were acquired at the beginning and end of each night. The whole data set of the LRa02 campaign consists of $836$ science frames for $231$ hours of observations. Magnitudes of the detected stars range from $11-18$ mag with a photometric precision of $1\%$ down to $15$ mag. Generally, observations were not performed during nights with either full Moon or strong wind ($>15m/s$). This was the case in $27$ nights during the campaign. Additional $20$ nights were lost due to technical problems during the final implementation phase of the robotic mode.         

\section{Photometric calibration of the scientific data}

The photometric calibration of the acquired data was performed with an automated data pipeline based on previous versions used for BEST (\cite{Rauer2004}) and modified to the BEST II system. The basic features of the pipeline are described below. A more detailed description of the data reduction process can be found in \cite{Kabath2007} and \cite{Karoff2007}.
  \begin{figure}
   \centering
   \includegraphics[angle=90, width=8cm, height=5cm]{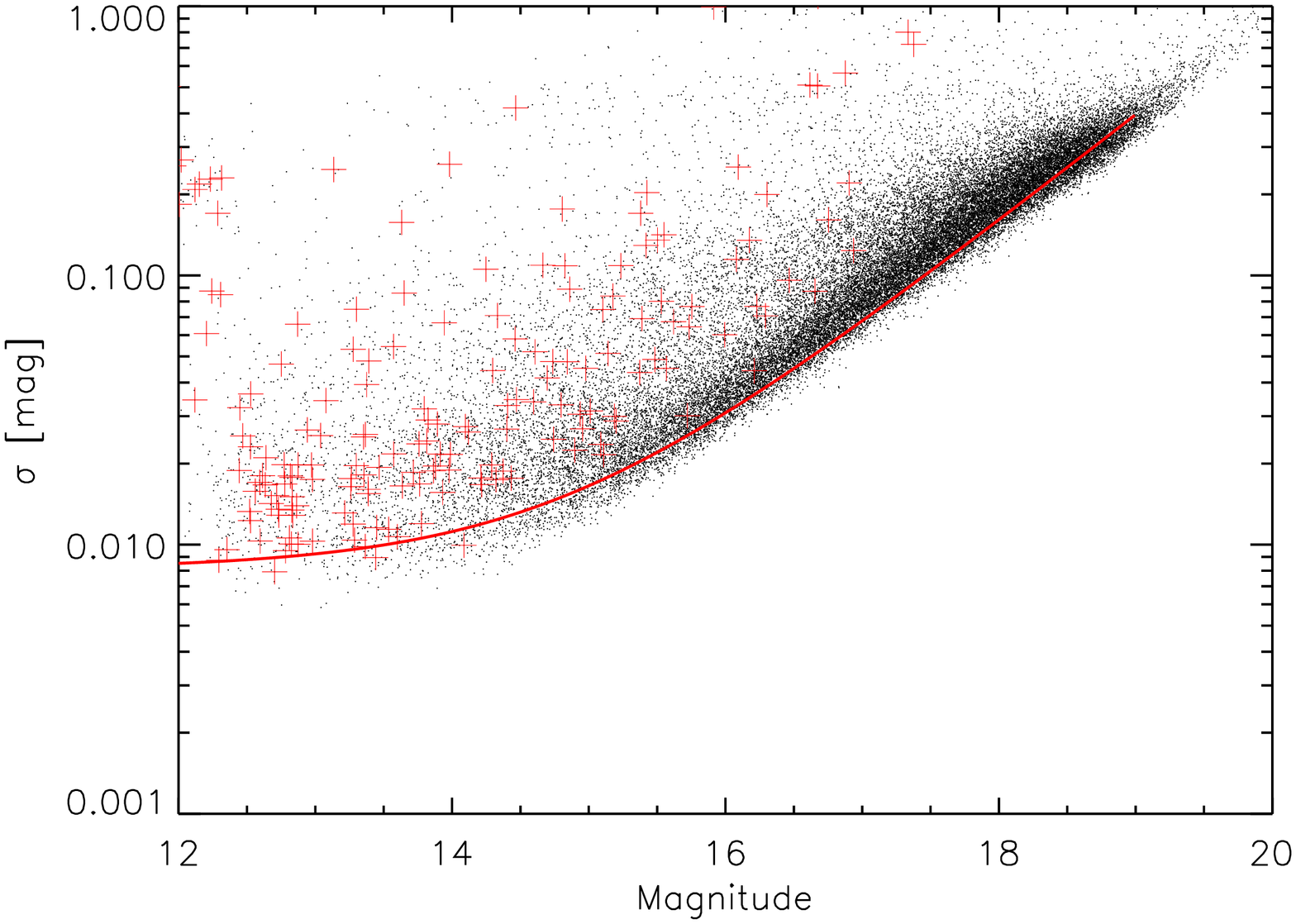}
   \includegraphics[angle=90, width=8cm, height=5cm]{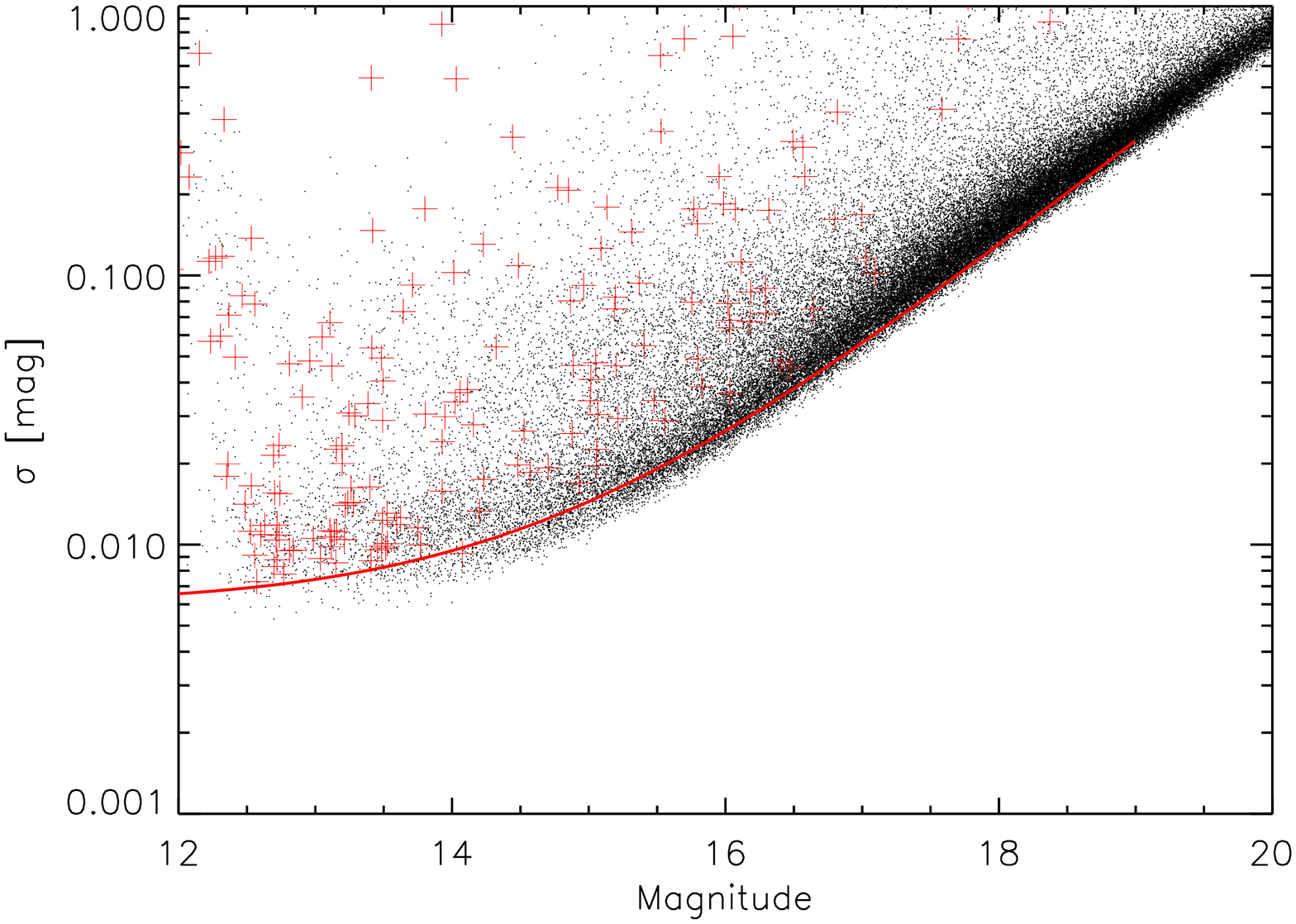}
      \caption{RMS plots for BEST II LRa02a (top) and LRa02b (bottom) over the whole observational campaign of $41$ nights. Red crosses represent detected periodic variable stars and the red solid line represents the limiting values of the RMS for the corresponding magnitudes.}
              \label{figrms}%
    \end{figure}
In a first step, the obtained scientific data were calibrated with bias, dark and flat field images taken each night during the observational sequence. These calibrated scientific images were then interpolated to a unified coordinate system. A reference frame for the transformation routine (\cite{Pal2007}) was first selected. Then, the transformation was applied on each image of the data set. In a second step, the data was processed using the image subtraction package ISIS (\cite{Alard1998}). Here a reference frame is selected and the image subtraction routine is subsequently applied to all images. For this purpose, a kernel function is defined on the reference frame and then subtracted from the PSF functions of each star on every frame. Thereafter, aperture photometry is applied to the image subtraction reference frame and in the subtracted images. Typically, the FWHM of a stellar point spread function is $4$ pixels. Therefore, a fixed aperture with a radius of $5$ pixels is used for the determination of the stellar flux and $15$ pixels for the determination of the background contribution. The obtained magnitudes were corrected for extinction. The remaining nightly offsets were corrected by using the $9000$ brightest and most constant stars for an estimate and then corrected on each image. In a third step, the internal coordinates were transformed to the world coordinate system using the USNO-A2.0 catalog and GRMATCH routine. The obtained transformation is applied to each star in the data set with the GRTRANS routine (\cite{Pal2007}). A subarcsecond precision is reached for the coordinate transformation. 

Finally, the BEST II relative magnitudes of a sample of several thousands constant stars is compared to the USNO-A2.0 catalog. Resulting deviations from the USNO-A2.0 match for constant stars are applied to each star from the data set. Due to accuracy limitations of the absolute magnitudes in the catalog used, the final BEST II absolute magnitudes have a slight uncertainty. However, since we are searching for differences in stellar light curves using differential photometry, this is of minor importance. In general, the error in the BEST II magnitude relative to an absolute photometric system after calibration with the USNO-A2.0 is usually not higher than $0.5$ mag. Moreover, stars from the BEST II field are cross-matched with the 2MASS catalog and the 2MASS ID as well.     

\section{Photometric quality of the data set}
 
The resulting light curves can still be differently affected by systematic effects during the nights. These effects may introduce false periodic signals that can be falsely identified as stellar variability. Therefore, a routine based on the SYSREM algorithm (\cite{Tamuz2005}) was applied to the whole data set to correct for possible systematic effects present at the same time in a significant number of stellar light curves. The algorithm works also if the true nature of the systematic effects remains unresolved. Figure~\ref{figrms} shows the final RMS plots for all stars observed over the whole campaign for both fields. Detected periodic variable stars are marked with red crosses. The solid line represents limiting RMS values for corresponding magnitudes. Under the excellent observational conditions at Cerro Armazones it is possible to reach a seeing of $0.66"$ (\cite{Rauer2008a}). The corresponding photometric precision of the measurements under such conditions is better than $1\%$ for more than $5000$ stars. During a typical photometric night at Cerro Armazones, it is possible to measure with a precision better than $1\%$ for up to $4000$ stars in the magnitude range 11-15 mag.

\section{Variability criteria}
\begin{figure}
   \centering
   \includegraphics[width=8cm, height=5cm]{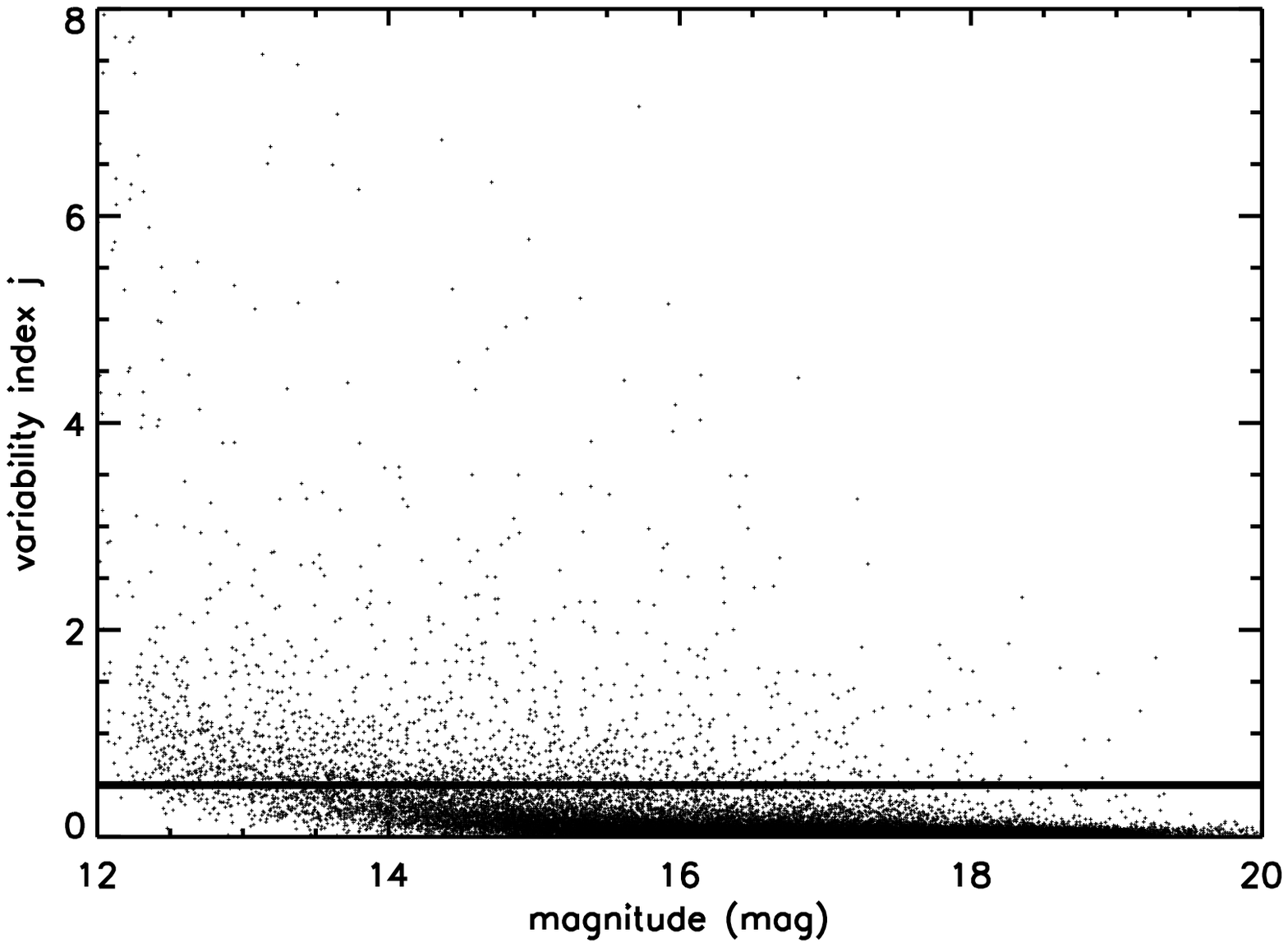}
   \includegraphics[width=8cm, height=5cm]{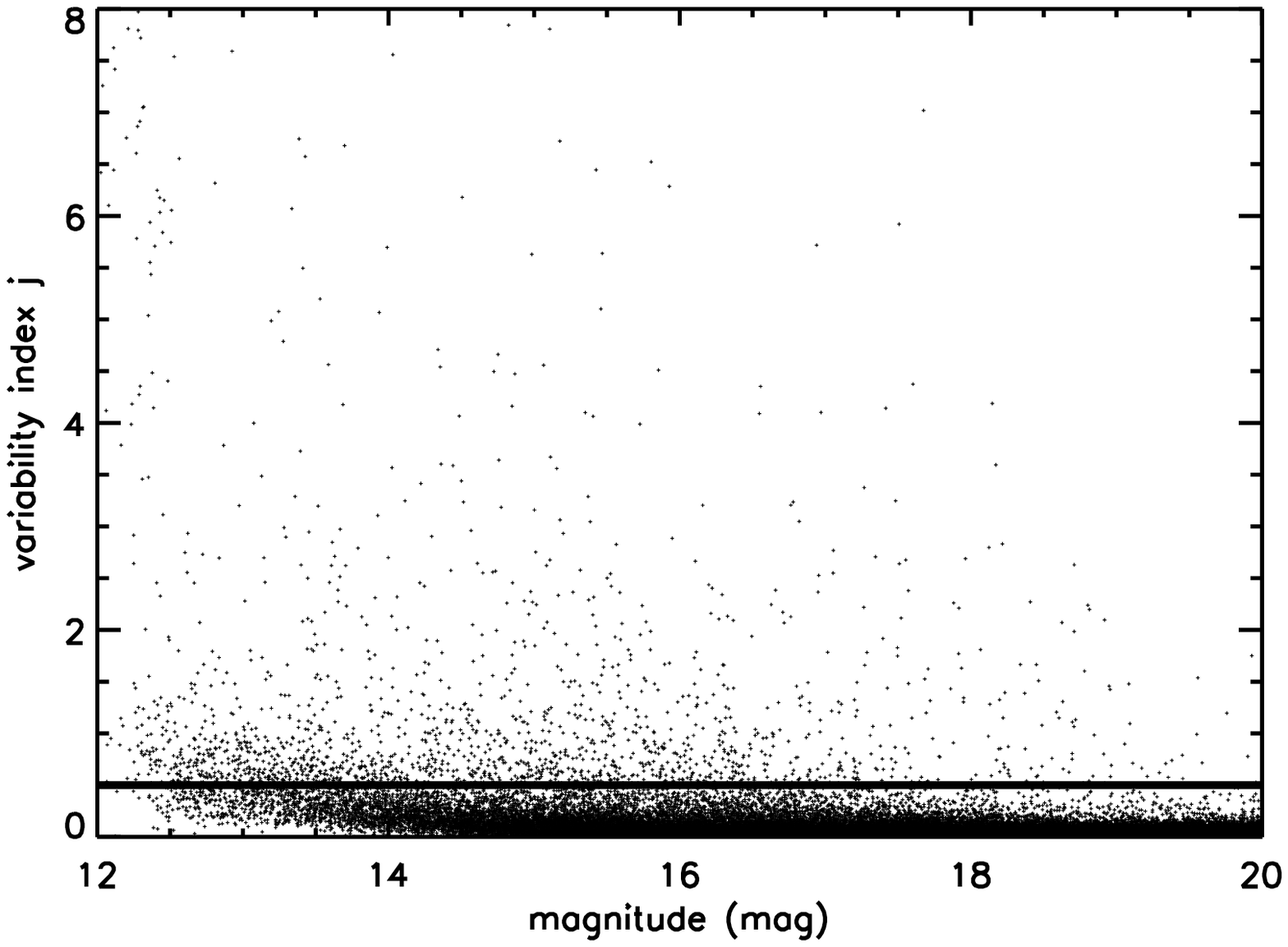}
   \caption{Distribution of variability index $j$ for stars in LRa02a (top) and LRa02b (bottom).}
              \label{figji}%
    \end{figure}
We detected $104335$ stars in both observed fields. Therefore, for the performed stellar variability characterization of the LRa02 field, an automatic routine that selects potentially variable stars from the data sample is needed. Thus, each star is marked with a $j$-index as defined by \cite{stetson}

\begin{equation}
j=\frac{\sum_{k=1}^{n}w_k sgn(P_{k}\sqrt{\left|P_{k}\right|})}{\sum_{k=1}^{n}w_k}
\end{equation}

\noindent where $k$ denotes consecutive pairs of observations, each with a weight $w_{k}$ and

\begin{equation}
P_{k} = \left\{ \begin{array}{ll}
\delta_{i(k)}\delta_{j(k)},\textrm{if}\; i(k) \neq j(k)\\
\delta_{i(k)}^2-1,\textrm{ if}\; i(k)=j(k)\end{array}\right.
\end{equation} 

\noindent where $\delta_{i(k)}$ and $\delta_{j(k)}$ are the residuals of normalized magnitudes from the mean magnitude of all data points for observations $i$ and $j$ within the pair $k$. The weighting factor was modified according to ~\cite{zhang} as

\begin{equation}
w_{k,i}=\exp{\frac{-\delta t_i}{\delta t}},
\end{equation}

\noindent where $t_i$ and $t$ is the time between two points and the time weighting interval, respectively. Figure~\ref{figji} presents the distribution of the variability index $j$ for all stars in both LRa02 data sets. A limiting value of $j=0.5$ was selected based on experience from previously characterized CoRoT fields (e.g. \cite{Kabath2009}) to mark potentially variable stars. In this case, $1858$ stars from LRa2a and $1868$ stars from LRa2b fields were marked for potential variability. However, to find the periodic variable stars as detected by BEST II, a further selection step must be performed.      

\section{Results}

Light curves of all potentially variable stars that were marked with $j>0.5$ were folded with the AoV algorithm (\cite{schwarzenberg-czerny}). Firstly, stellar light curves were folded within a period range of $0.1$ to $35$ days. Secondly, these light curves were further inspected visually. Stars with periods close to $1$ day or multiples thereof were rejected if they did not show clear natural variability. Artificial variability is likely due to the diurnal cycle or changes of the background level, temperature or airmass. We found \varisa newly detected periodic variable stars in LRa2a and \varisb in LRa2b. The number of detected periodic variable stars with BEST II is directly comparable with detection rates of ASAS 3 (\cite{Pojmanski2002}) and OGLE-II (\cite{Udalski1997}) surveys as presented in Table 1. In the case of OGLE-II it is not specified if the number of variable stars concerns only periodic variable stars or also irregular objects. An absolute number of detected stars with each survey is also shown. 

\begin{table}
\caption{Detection rates for variable stars of BEST II and surveys with values based on~\cite{Eyer}.}             
\label{tabsumm2}      
\centering                          
\begin{tabular}{l l l l}        
\hline\hline                 
Property      &  BEST II LRa02& ASAS 3& OGLE II\\\hline
Stars detected &  104,335 & 7,300,000& 30,000,000\\
Variable stars& \variables ($0.3\%$)& 24,936 ($0.3\%$)& 200,000 ($0.67\%$) \\
\hline                                   
\end{tabular}
\end{table}

\begin{figure}
   \centering
   \includegraphics[width=.49\textwidth]{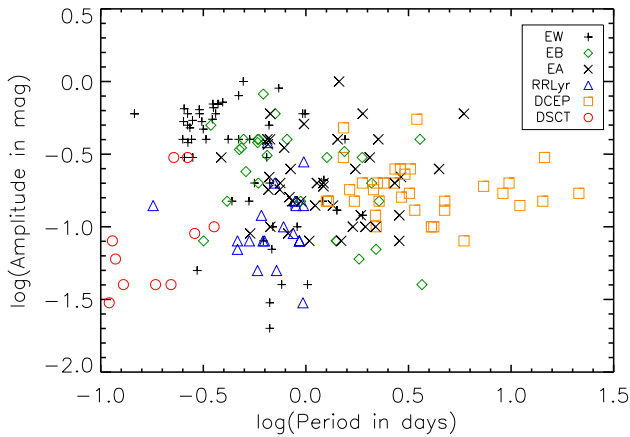}
      \caption{Different classification types of variable stars found in LRa02 in a logarithmic plot of amplitude versus period. Each type of symbol corresponds to one of EA, EB, EW, DSCT, DCEP, RRLyr classes.}
              \label{varstars}%
    \end{figure}

Newly detected variable stars are classified with a GCVS-based reduced schema (see ~\cite{sterken}). Here, stars are divided into two groups of stellar variability. Pulsating stars are described by types DCEP ($\delta$ Cephei prototype pulsating stars), DSCT ($\delta$ Sct type pulsating stars), RR Lyr (if possible also subtypes RRa, RRb, RRc), SXPHE (prototype SX Phe), $\gamma$ Dor (prototype $\gamma$ Doradus). Furthermore, a PULS class is introduced where the type could not be classified uniquely. Eclipsing stars were subdivided into EA (Algol type eclipsing stars), EB ($\beta$ Lyr type eclipsing stars) and EW (W UMa type eclipsing stars) classes. An additional group of stars showing variability characteristic of spotted (stellar spots) or elliptical stars was classified as ELL and/or SP. Finally, some stars whose light curve variation is due to their rotation and magnetic field were classified as $\alpha^{2}$CVn. A limited number of stars was classified as cataclysmic variables or as VAR, in particular those where more data is needed. Our classification scheme is illustrated in Figure~\ref{varstars}, where a logarithmic plot of amplitude against period of detected variable stars is shown. The classes EA, EB, EW, DSCT, DCEP, RRLyr are represented with different symbols. Detailed statistics on the types of newly found variable stars are presented in Table 3. A detailed information on identified stars is shown Table 4 and Table 5. 

Since the BEST II data set possesses no color information, periodic variable stars were cross-identified with the 2MASS catalog \cite{Skrutskie2006}. Information about the 2MASS ID is inserted in Table 3 and Table 4. 2MASS magnitudes and IDs provide extended spectral information and allow us to cross-match with other catalogs.


\subsection{Pulsating variable stars}

Pulsating stars with periods below $0.3$ days were classified as DSCT. However, a few stars with periods of about $0.1$ days were classified as SXPHE due to the characteristic light curve and period of the prototype star SX Phoenicis. For the DSCT with periods below $0.3$ days, the classification should be fairly unique. Furthermore, some of the DSCT stars such as no. $lra2b\_00190$ and $lra2b\_01387$ show clear multiperiodicity. Thus additional frequency analysis of the light curves would be needed. A few stars within a period range of $0.3$ to $5$ days, with low amplitudes $<0.1$ mag and characteristic light curve shape were characterized as $\gamma$Dor (\cite{kaye1999}). Stars with periods longer than $1$ day and amplitudes $>0.1$ magnitudes were classified as $\delta$CEP stars. The RR Lyr stars were classified based on the characteristic shape of the light curve and period range between $0.3$ and $1$ day. 

\subsection{Eclipsing variable stars}

Eclipsing stars of the EA class show nearly constant light curves between eclipses. They can show the so-called reflection effect, causing some small modulation of the light curve (\cite{Vaz1985}, \cite{Pariah1998}), in particular no. $lra2a00492$ and $lra2b00821$. Stars belonging to the EB class vary continuously between eclipses and stars from the EW class have generally near equal depth and periods shorter than $1$ day. Several eclipsing binaries, primarily some EW, show the O'Connell effect, i.e. a deformation of the light curve due to the presence of stellar spots (\cite{Oconnel}). The effect is evident in particular in light curves no. $lra2a00302$, $lra2a\_00343$, $lra2a\_00621$, $lra2a\_01627$ and $lra2b\_01334$.

The classification of rotating and spotted stars ELL/SP is based on unequal minimum/maximum depth of the light curves. Some of the systems classified as spotted also may be cataclysmic variable stars in the quiescent phase. Further color and spectroscopic information is needed to clarify this result.

Stars showing a slight amplitude variability in the region of a few percent which do not fulfill the classification criteria for any of the abovementioned variability types were classified as $\alpha^{2}CVn$. The prototype is Cor Caroli from the constellation Canum Venaticorum, and the characteristics are usually fast rotation and the presence of emission lines in the CaII H, K and H$_{\alpha}$ bands, indicating strong chromospheric activity (\cite{Fernandez1994}).

\subsection{Previously known variable stars in the CoRoT LRa02 field}

Detected variable stars from the BEST II data set have been cross-matched with the SIMBAD\footnote{http://simbad.u-strasbg.fr/simbad/} and AAVSO VSX\footnote{http://www.aavso.org/vsx} catalogs. Comments on light curves of already known variable stars reobserved with BEST II are provided in Table 1. In general, previously reported variable stars that are brighter than $11$ mag are saturated in the BEST II data. We were able to confirm only short periodic variable stars due to our duty cycle. The following stars were found in BEST II data sets: V$0452$\nolinebreak[4] Mon, V$0376$\nolinebreak[4] Mon which is also ASAS\nolinebreak[4] $064848-0336.3$, XZ\nolinebreak[4] Mon,\linebreak[4] ASAS\nolinebreak[4] $064750-0352.8$ (\cite{Pojmanski2002}) and EI\nolinebreak[4] Mon (\cite{Ahnert1949}). The relevant stars from the BEST\nolinebreak[4] II set are marked with their previous catalog names in Table 3 and Table 4.

\begin{table}
\caption{Previously known variable stars in LRa02 cross-matched with BEST II data set.}             
\label{tabr}      
\centering
\renewcommand{\footnoterule}{}                          
\begin{tabular}{l c l }        
\hline\hline                 
Star ID      &  BEST II ID & Comments \\
\hline
	XZ Mon 						& $lra2a\_01702$	& confirmed\\
	EI Mon						& $lra2b\_00469$	& confirmed\\
	V$0376$ Mon					& $lra2a\_00311$	& P=1.6521 days confirmed\\
	V$0452$ Mon 				& $lra2a\_00788$	& confirmed\\
	ASAS $064750$-$0352.8$	& $lra2a\_00134$	& confirmed\\
\hline                                   
\end{tabular}
\end{table}

\section{Summary} 
We observed CoRoT's LRa02 stellar field with the BEST II telescope during $41$ nights from November $2007$ to March $2008$. The obtained data were searched for periodic variable stars to support CoRoT's additional science programs. In a sample of \textbf{$104335$} light curves, we identified \variables stars showing regular periodicity. Almost all of these periodic variable stars are new detections by BEST II with periods ranging from $0.1$ to $35$ days. A classification of the periodic variables has been performed and the members of the resolved classes are specified. In addition, we confirmed variability for $5$ already known stars in the observed field. More detailed information and finding charts will be provided upon request.

\begin{table}
\caption{Statistics on newly detected variable stars with BEST II in the LRa02 stellar field compared with OGLE.}             
\label{tabsumm}      
\centering                          
\begin{tabular}{l c l }        
\hline\hline                 
Property      &  LRa02a & LRa02b \\\hline
Nr. of stars detected 	&  37361   			& 66974\\
Stars with $j>0.5$		& 1858 ($5\%$)		& 1868 ($3\%$)\\
Periodic variable stars	& \varisa ($0.5\%$)	& \varisb ($0.3\%$)\\
   DSCT 				&4  	&3\\
   DCEP				&13   &24\\
   RR Lyr 			&12 	&6\\
   $\alpha^2$CVn	&16   &13\\
   EA(Algol) 		&27 	&20\\
   EB($\beta$ Lyr)&16 	&15\\
   EW (W UMa)		&19	&29\\
   other				&66 	&67\\
\hline 
\end{tabular}
\end{table}

\begin{acknowledgements}
This work was funded by Deutsches Zentrum f\"{u}r Luft- und Raumfahrt and partly by the Nordrhein-Westf\"{a}lische Akademie der Wissenschaften. The authors gratefully acknowledge the support and assistance of the administration of the Universidad Cat\'{o}lica del Norte (UCN) in Antofagasta, Chile. The great support and help from the UCN staff based at OCA is also appreciated. The authors would like to thank Hartmut Korsitzky, Harald Michaelis, Andreas Kotz and Christopher Carl, who worked on the design and installation of the telescope system, and to Martin Paegert and Holger Drass who supported the project in the maintenance and adjusting phase. We made use of SIMBAD, 2MASS, GCVS catalogues and AAVSO variable star search index. P. K. acknowledges partial support covering the fee for the first CoRoT Symposium held in Paris in February 2009. We are also most grateful to the anonymous referee for helpful comments and useful advice.
\end{acknowledgements}

\newpage

\renewcommand{\footnoterule}{}
\longtabL{4}{
\begin{landscape}
\small{
\begin{longtable}{lcllccccc}
\caption{\label{tabres1} Periodic variable stars detected in LRa2a.}\\
\hline\hline
BEST ID &2MASS ID &$\alpha$(J2000) & $\delta$(J2000) & Mean mag(mag) & Period (days) & Amplitude (mag) & Type & Other Names\\
\hline
\endfirsthead
\caption{continued.}\\
\hline\hline
BEST ID &2MASS ID &$\alpha$(J2000) & $\delta$(J2000) & Mean mag(mag) & Period (days) & Amplitude (mag) & Type & Other Names\\
\hline
\endhead
\hline
\endfoot
lra2a\_00004*\footnote{Stars with asterisk are located in the CoRoT's exoplanetary field-of-view.}& 	06472576-0354198&      6      47    25.8&      -3      54  20.4&   15.172&    0.682&  0.07&EW &\\
lra2a\_00074*& 06475959-0335092&      6      47    59.6&      -3      35   9.4&   12.937&    0.989&  0.04&SP &\\
lra2a\_00079*& 06480297-0333011 &     6      48     3.0&      -3      33   1.2&   12.703&    1.024&  0.06&$\gamma$Dor &\\
lra2a\_00092& 	06473486-0357256  &    6      47    34.9&      -3      57  26.1&   15.166&    1.402&  0.08&EB &\\
lra2a\_00133*& 06481854-0329035  &    6      48    18.6&      -3      29   3.7&   15.456&    2.186&  0.10&EA &\\
lra2a\_00134& 06474958-0352496&      6      47    49.6&      -3      52  50.0&   12.344&    2.310&  0.20&EA  &ASAS 064750-0352.8\\
lra2a\_00138*& 06481445-0333074  &    6      48    14.5&      -3      33   7.7&   15.398&    3.449&  0.10&SP &\\
lra2a\_00143& 	06474412-0359120  &    6      47    44.1&      -3      59  12.4&   14.218&    1.039&  0.08&EA &\\
lra2a\_00159& 06474131-0404558  &    6      47    41.3&      -4       4  56.1&   16.121&    0.436&  0.15&EW &\\
lra2a\_00186& 06474594-0405436  &    6      47    46.0&      -4       5  42.0&   15.098&    0.180&  0.14&RRcLyr &\\
lra2a\_00194*& 	06484004-0324130  &    6      48    40.0&      -3      24  13.2&   15.631&    1.950&  0.10&EA &\\
lra2a\_00203& 06473681-0417436  &    6      47    36.8&      -4      17  43.8&   15.899&   20.470&  0.20&VAR &\\
lra2a\_00221*& 06482041-0345317  &    6      48    20.4&      -3      45  31.9&   15.913&    0.663&  0.22&EA &\\
lra2a\_00223*& 06484348-0327173  &    6      48    43.5&      -3      27  17.3&   17.279&    0.649&  0.40&EA &\\
lra2a\_00238& 06475752-0407048  &    6      47    57.5&      -4       7   5.1&   14.618&    6.398&  0.19&SP &\\
lra2a\_00244*&  06480847-0359231 &    6      48     8.5&      -3      59  23.3&   12.308&    3.678&  0.04&EB &\\
lra2a\_00245&  06472906-0431527 &    6      47    29.0&      -4      31  52.2&   12.317&    0.666&  0.20&EW &\\
lra2a\_00259*& 06485379-0325069  &    6      48    53.8&      -3      25   7.2&   14.927&    0.541&  0.08&VAR &\\
lra2a\_00269& 	06473121-0432567  &    6      47    31.1&      -4      32  56.2&   15.044&    2.909&  0.22&EA &\\
lra2a\_00295& 06473586-0432542  &    6      47    35.8&      -4      32  53.7&   16.204&    0.8425&  0.25&EA &\\
lra2a\_00298*& 06491310-0314030  &    6      49    13.1&      -3      14   2.8&   14.904&   21.388&  0.15&$\alpha^2$CVn &\\
lra2a\_00301*& 06485712-0327249  &    6      48    57.1&      -3      27  25.2&   14.538&    2.927&  0.12&DCEP &\\
lra2a\_00302*& 06485993-0325275  &    6      48    59.9&      -3      25  27.7&   15.764&    0.510&  0.24&EB &\\
lra2a\_00304*& 06484317-0339351  &    6      48    43.2&      -3      39  35.3&   16.911&    0.350&  0.55&EW &\\
lra2a\_00311& 06484802-0336168  &    6      48    48.0&      -3      36  16.8&   12.564&    1.652&  0.45& DCEP &V0376 Mon\\  
lra2a\_00319& 06474707-0427025  &    6      47    47.1&      -4      27   2.4&   14.440&    0.955&  0.06&SP &\\
lra2a\_00325*& 06484192-0343065  &    6      48    41.9&      -3      43   6.6&   15.860&   11.773&  0.00&LP &\\
lra2a\_00342*& 06480352-0416460  &    6      48     3.5&      -4      16  46.2&   15.741&    3.474&  0.55&DCEP &\\
lra2a\_00343*& 06492660-0309246  &    6      49    26.6&      -3       9  24.8&   14.884&    0.474&  0.34&EB &\\
lra2a\_00346*& 06482771-0357328  &    6      48    27.7&      -3      57  33.1&   12.631&    2.375&  0.05&ELL &\\
lra2a\_00356*& 06490979-0325538  &    6      49     9.8&      -3      25  54.2&   14.181&    1.543&  0.06&ELL/SP &\\
lra2a\_00408*&  06482880-0409053 &    6      48    28.8&      -4       9   5.6&   15.752&    0.868&  0.10&SP &\\
lra2a\_00416*& 06483188-0407577  &    6      48    31.9&      -4       7  57.9&   16.611&    0.750&  0.20&EA &\\
lra2a\_00435*c\footnote{Potential stellar crowding affecting the light curve of stars is marked with $c$.}& 06492145-0332384  &    6      49    21.5&      -3      32  39.2&   16.091&    0.265&  0.50&EW &\\
lra2a\_00438*c& 06492144-0332442  &    6      49    21.4&      -3      32  44.3&   16.014&    0.265&  0.60&EW &\\
lra2a\_00450*c& 06491193-0342336  &    6      49    11.9&      -3      42  33.7&   16.162&    0.272&  0.48&EW &\\
lra2a\_00459*& 06481562-0430030  &    6      48    15.6&      -4      30   3.1&   16.825&    0.784&  0.35&EA &\\
lra2a\_00466*& 06490094-0354075  &    6      49     0.9&      -3      54   7.6&   12.957&   12.878&  0.15&ELL &\\
lra2a\_00476*& 06485459-0401340  &    6      48    54.6&      -4       1  33.8&   14.817&    0.565&  0.06&VAR &\\
lra2a\_00477*& 06482272-0427312  &    6      48    22.7&      -4      27  31.4&   14.601&   10.863&  0.10&VAR &\\
lra2a\_00479*& 06485491-0401291  &    6      48    54.9&      -4       1  29.2&   14.523&    0.565&  0.06&VAR &\\
lra2a\_00485*& 06485905-0359090  &    6      48    59.1&      -3      59   9.1&   15.334&    3.964&  0.11&VAR &\\
lra2a\_00488*& 06481978-0432424  &    6      48    19.8&      -4      32  42.4&   14.270&    0.266&  0.38&EW &\\
lra2a\_00491*& 06482534-0428360  &    6      48    25.3&      -4      28  36.2&   13.542&    0.220&  0.04&DSCT &\\
lra2a\_00492*& 06495531-0315442  &    6      49    55.3&      -3      15  44.4&   15.468&    0.653&  0.37&EA &\\
lra2a\_00495*& 06484754-0411406  &    6      48    47.6&      -4      11  40.7&   14.067&    2.875&  0.08&$\alpha^2$CVn &\\
lra2a\_00517*& 06483463-0425378  &    6      48    34.6&      -4      25  38.1&   14.936&    2.890&  0.18&VAR &\\
lra2a\_00524*& 06484810-0416130  &    6      48    48.1&      -4      16  13.3&   13.309&   33.768&  0.00&LP &\\
lra2a\_00529*& 06492898-0343193  &    6      49    29.0&      -3      43  19.4&   14.924&    5.533&  0.08&$\alpha^2$CVn &\\
lra2a\_00531*& 06500331-0315474  &    6      50     3.3&      -3      15  47.5&   16.466&    0.674&  0.30&EA &\\
lra2a\_00533*& 06484980-0416279  &    6      48    49.8&      -4      16  28.0&   13.193&   35.176&  0.17&LP &\\
lra2a\_00549*& 06493894-0339172  &    6      49    38.9&      -3      39  17.3&   12.738&    0.055&  0.03&DSCT &\\
lra2a\_00551*& 06492376-0351520  &    6      49    23.8&      -3      51  52.0&   14.829&    3.735&  0.04&$\alpha^2$CVn &\\
lra2a\_00559*& 06490729-0406270  &    6      49     7.3&      -4       6  27.2&   13.936&    1.780&  0.09&ELL &\\
lra2a\_00571*& 06491929-0358586  &    6      49    19.3&      -3      58  58.6&   16.814&   24.413&  0.20&VAR &\\
lra2a\_00579*& 06494398-0340111  &    6      49    44.0&      -3      40  11.3&   12.369&    1.018&  0.04&EW &\\
lra2a\_00586*& 06491682-0403076  &    6      49    16.8&      -4       3   7.8&   13.180&    0.990&  0.01&VAR &\\
lra2a\_00601*& 06485506-0423379  &    6      48    55.1&      -4      23  38.2&   15.337&    0.976&  0.28&RRLyr &\\
lra2a\_00603*& 06483611-0440018  &    6      48    36.1&      -4      40   2.0&   13.204&    1.024&  0.01&$\gamma$Dor &\\
lra2a\_00621*& 06501378-0323200  &    6      50    13.8&      -3      23  20.4&   14.380&    0.809&  0.40&EB &\\
lra2a\_00637*& 06491871-0410214  &    6      49    18.7&      -4      10  21.6&   16.482&    2.780&  0.20&DCEP &\\
lra2a\_00642*& 06500981-0329026  &    6      50     9.8&      -3      29   2.7&   14.562&    4.226&  0.09&ELL/SP &\\
lra2a\_00646*& 06495951-0337371  &    6      49    59.5&      -3      37  37.2&   14.799&    2.741&  0.04&ELL/SP &\\
lra2a\_00649*& 06492236-0408156  &    6      49    22.4&      -4       8  15.8&   12.596&    0.581&  0.05&RRLyr &\\
lra2a\_00652*& 06494921-0346398  &    6      49    49.2&      -3      46  40.0&   13.579&    0.185&  0.04&SXPHE/DSCT &\\
lra2a\_00656*& 06493211-0401410  &    6      49    32.1&      -4       1  41.1&   14.579&   14.616&  0.06&$\alpha^2$CVn &\\
lra2a\_00662*& 06502316-0321053  &    6      50    23.2&      -3      21   5.7&   14.607&    0.820&  0.09&EA &\\
lra2a\_00663*& 06495275-0346122  &    6      49    52.7&      -3      46  12.4&   14.219&   16.178&  0.04&LP &\\
lra2a\_00673*& 06490980-0424545  &    6      49     9.8&      -4      24  54.6&   13.997&    1.893&  0.60&EA &\\
lra2a\_00683*&06492634-0413091   &    6      49    26.3&      -4      13   9.3&   13.181&    0.317&  0.08&EB &\\
lra2a\_00721*& 06504255-0319206  &    6      50    42.6&      -3      19  21.0&   12.861&    0.905&  0.11&ELL/SP &\\
lra2a\_00732*&06494128-0411169   &    6      49    41.3&      -4      11  16.9&   14.161&    3.074&  0.07&ELL/SP &\\
lra2a\_00742*&06504787-0318096   &    6      50    47.9&      -3      18  10.1&   11.962&    1.448&  0.07&VAR &\\
lra2a\_00744*&06493211-0419598   &    6      49    32.1&      -4      19  59.9&   15.341&   33.507&  0.12&LP &\\
lra2a\_00757*&06493592-0418296   &    6      49    35.9&      -4      18  29.7&   15.391&    0.535&  0.09&EA &\\
lra2a\_00761*&06495752-0401401   &    6      49    57.5&      -4       1  40.0&   14.707&    2.201&  0.10&DCEP &\\
lra2a\_00764*&06500411-0357205   &    6      50     4.1&      -3      57  20.5&   13.508&    1.012&  0.03&$\gamma$Dor &\\
lra2a\_00771*&06491012-0442323   &    6      49    10.1&      -4      42  32.5&   16.383&   28.418&  0.00&LP &\\
lra2a\_00788*&06495594-0410207   &    6      49    55.9&      -4      10  20.7&   12.630&    1.539&  0.20& EA &V0452 Mon\\
lra2a\_00799*&06494455-0421312   &    6      49    44.6&      -4      21  31.4&   14.938&    4.475&  0.04&$\alpha^2$CVn &\\
lra2a\_00808*& 06510307-0319432  &    6      51     3.1&      -3      19  43.6&   15.582&    0.868&  0.10&VAR &\\
lra2a\_00809*&06503829-0340019   &    6      50    38.3&      -3      40   2.0&   13.759&   31.362&  0.00&LP &\\
lra2a\_00811*&06503672-0341301   &    6      50    36.7&      -3      41  30.3&   13.226&    1.816&  0.06&EB &\\
lra2a\_00817*&06492766-0439506   &    6      49    27.7&      -4      39  50.7&   12.093&    0.357&  0.10&DSCT &\\
lra2a\_00820*&06501761-0359301   &    6      50    17.6&      -3      59  29.9&   15.028&    0.669&  0.40&EA &\\
lra2a\_00822*&06495696-0416327   &    6      49    57.0&      -4      16  32.7&   14.880&    0.886&  0.06&VAR &\\
lra2a\_00839*& 06500179-0414283  &    6      50     1.8&      -4      14  28.3&   13.663&    7.326&  0.04&$\alpha^2$CVn &\\
lra2a\_00851*&06502424-0358373   &    6      50    24.2&      -3      58  37.1&   16.702&    1.742&  0.25&EA &\\
lra2a\_00855*c&06505170-0336412   &    6      50    51.7&      -3      36  41.3&   15.114&    0.929&  0.08&RRLyr &\\
lra2a\_00856*c&06505181-0336367   &    6      50    51.8&      -3      36  37.6&   15.109&    0.929&  0.08&RRLyr &\\
lra2a\_00862*&06503880-0348144   &    6      50    38.8&      -3      48  14.5&   13.776&    3.709&  0.12&EA &\\
lra2a\_00863*&06500327-0417088   &    6      50     3.3&      -4      17   8.8&   13.287&    0.967&  0.03&RRLyr &\\
lra2a\_00873*&	06495873-0422291   &    6      49    58.7&      -4      22  29.0&   15.441&    0.607&  0.12&RRLyr &\\
lra2a\_00876*&06501226-0411535   &    6      50    12.2&      -4      11  53.3&   13.892&    0.666&  0.04&VAR &\\
lra2a\_00879*c&06511257-0323040   &    6      51    12.6&      -3      23   4.7&   15.722&    1.416&  0.13&EW &\\
lra2a\_00880*c&	06511248-0323088   &    6      51    12.5&      -3      23   8.9&   15.723&    1.416&  0.13&EW &\\
lra2a\_00891*&06510465-0330432   &    6      51     4.7&      -3      30  43.4&   16.664&    0.588&  0.40&EB &\\
lra2a\_00896*& 06500785-0418300  &    6      50     7.8&      -4      18  30.1&   17.238&    0.979&  0.51&EA &\\
lra2a\_00900*& 06510525-0332513  &    6      51     5.3&      -3      32  51.7&   12.970&    5.573&  0.08&$\alpha^2$CVn &\\
lra2a\_00925*&06501489-0417160   &    6      50    14.9&      -4      17  16.0&   13.680&    1.672&  0.06&VAR &\\
lra2a\_00926*&06500533-0425178   &    6      50     5.3&      -4      25  17.8&   14.211&   13.305&  0.06&VAR &\\
lra2a\_00940*&06511871-0327416   &    6      51    18.7&      -3      27  41.7&   16.507&    4.460&  0.25&EA &\\
lra2a\_00950*&06501612-0420369   &    6      50    16.1&      -4      20  36.9&   16.140&    9.123&  0.17&DCEP &\\
lra2a\_00963*&	06501580-0421518  &    6      50    15.8&      -4      21  51.7&   13.614&   21.337&  0.03&LP &\\
lra2a\_00968*&	06495042-0443181  &    6      49    50.4&      -4      43  18.0&   13.485&    9.906&  0.06&LP &\\
lra2a\_01071*& 06510707-0359092  &    6      51     7.1&      -3      59   9.2&   16.239&    0.316&  0.47&EW &\\
lra2a\_01088*& 06505503-0411143  &    6      50    55.0&      -4      11  14.3&   13.948&    4.752&  0.06&$\alpha^2$CVn &\\
lra2a\_01098*& 06514793-0331527  &    6      51    47.9&      -3      31  52.9&   14.504&    1.355&  0.14&EA &\\
lra2a\_01103*& 06512209-0354038  &    6      51    22.1&      -3      54   3.9&   14.005&    0.393&  0.72&EW &\\
lra2a\_01111*& 06503162-0436431  &    6      50    31.6&      -4      36  43.4&   16.200&    2.137&  0.13&VAR &\\
lra2a\_01123*& 	06511923-0359522  &    6      51    19.2&      -3      59  52.2&   13.974&    4.202&  0.10&DCEP &\\
lra2a\_01125*& 06511002-0407400  &    6      51    10.0&      -4       7  39.9&   18.003&    1.452&  1.00&EA &\\
lra2a\_01126*& 	06520050-0326327  &    6      52     0.5&      -3      26  32.9&   13.676&    2.711&  0.20&EA &\\
lra2a\_01127*& 06520077-0326255  &    6      52     0.8&      -3      26  25.8&   14.311&    1.694&  0.10&EA &\\
lra2a\_01137*& 06514038-0344546  &    6      51    40.4&      -3      44  54.9&   14.954&    8.149&  0.08&$\alpha^2$CVn &\\
lra2a\_01163*& 06511819-0407406  &    6      51    18.2&      -4       7  40.4&   15.243&    0.530&  0.08&RRLyr &\\
lra2a\_01168*& 06515875-0335435  &    6      51    58.8&      -3      35  44.0&   12.122&    0.665&  0.20&VAR &\\
lra2a\_01169*& 06515256-0340546  &    6      51    52.6&      -3      40  54.9&   15.197&    3.941&  0.10&VAR &\\
lra2a\_01170*& 06514139-0349581  &    6      51    41.4&      -3      49  58.2&   12.295&    1.010&  0.02&$\gamma$Dor &\\
lra2a\_01184*& 06512594-0406260  &    6      51    25.9&      -4       6  25.8&   15.992&    1.723&  0.15&VAR &\\
lra2a\_01191*& 06515385-0345415  &    6      51    53.9&      -3      45  41.9&   12.580&    0.666&  0.02&EW &\\
lra2a\_01194c*& 06503821-0447288 &     6      50    38.2&      -4      47  29.0&   15.913&   32.136&  0.00&LP/CV &\\
lra2a\_01196c*& 	06503844-0447354 &     6      50    38.4&      -4      47  35.6&   15.777&   32.159&  0.00&LP/CV &\\
lra2a\_01211*& 	06520334-0342160  &    6      52     3.4&      -3      42  16.3&   17.281&    0.620&  0.82&EB &\\
lra2a\_01212*& 06511753-0420239  &    6      51    17.5&      -4      20  23.8&   13.538&    0.762&  0.04&EW &\\
lra2a\_01219*& 	06514910-0356440  &    6      51    49.1&      -3      56  44.1&   12.680&    0.665&  0.03&EW &\\
lra2a\_01224*& 	06521674-0335553  &    6      52    16.8&      -3      35  55.6&   13.772&    1.881&  0.30&EB &\\
lra2a\_01245*& 06521143-0343311  &    6      52    11.4&      -3      43  31.4&   13.396&    1.700&  0.10&VAR &\\
lra2a\_01246*&  06514341-0406202 &    6      51    43.4&      -4       6  20.2&   16.212&    2.271&  0.15&EB &\\
lra2a\_01253*&  	06505580-0446098 &    6      50    55.8&      -4      46   9.9&   13.632&    6.817&  0.06&ELL/SP &\\
lra2a\_01265*& 06512426-0425509   &    6      51    24.3&      -4      25  50.8&   15.637&    3.394&  0.13&DCEP &\\
lra2a\_01269*& 06521408-0346014  &    6      52    14.1&      -3      46   1.7&   12.845&    0.628&  0.08&RRLyr &\\
lra2a\_01293*& 	06522569-0340401  &    6      52    25.7&      -3      40  40.4&   15.697&    0.704&  0.20&RRLyr &\\
lra2a\_01297*& 06513157-0425357  &    6      51    31.6&      -4      25  35.7&   13.383&    0.846&  0.14&EW &\\
lra2a\_01298*& 	06521428-0351005  &    6      52    14.3&      -3      51   0.3&   16.919&    0.650&  0.38&RRLyr &\\
lra2a\_01310*& 06522283-0347013  &    6      52    22.8&      -3      47   1.5&   14.390&    5.341&  0.05&VAR &\\
lra2a\_01314*& 06524208-0332543  &    6      52    42.1&      -3      32  54.5&   14.454&    0.661&  0.50&EB &\\
lra2a\_01338&  06524845-0332472 &    6      52    48.4&      -3      32  47.3&   13.972&    5.896&  0.08&DCEP &\\
lra2a\_01358*& 	06514548-0429172  &    6      51    45.5&      -4      29  17.0&   14.570&   14.277&  0.15&DCEP &\\
lra2a\_01366& 06525102-0336570  &    6      52    51.0&      -3      36  57.2&   14.840&    9.298&  0.04&$\alpha^2$CVn &\\
lra2a\_01372*& 06522231-0401110  &    6      52    22.3&      -4       1  11.2&   15.020&    0.485&  0.35&EB &\\
lra2a\_01381& 06530018-0332288  &    6      53     0.2&      -3      32  28.8&   16.424&    0.583&  0.40&EB &\\
lra2a\_01384*& 06523144-0356300  &    6      52    31.4&      -3      56  30.2&   14.003&    1.083&  0.04&$\gamma$Dor &\\
lra2a\_01394& 06524198-0350201  &    6      52    42.0&      -3      50  20.4&   15.497&    6.133&  0.08&$\alpha^2$CVn &\\
lra2a\_01396& 06530279-0333359  &    6      53     2.8&      -3      33  35.8&   14.063&    2.754&  0.05&VAR &\\
lra2a\_01405*& 06522763-0404042  &    6      52    27.6&      -4       4   4.2&   14.623&   14.576&  0.05&VAR &\\
lra2a\_01427*& 06520097-0428571  &    6      52     1.0&      -4      28  57.0&   15.200&    1.589&  0.06&ELL/SP &\\
lra2a\_01429*& 06515040-0437579  &    6      51    50.4&      -4      37  57.8&   14.641&   17.165&  0.08&$\alpha^2$CVn &\\
lra2a\_01441& 	06524826-0355267  &    6      52    48.3&      -3      55  26.9&   14.253&    0.868&  0.09&RRLyr &\\
lra2a\_01456*& 06515958-0438103  &    6      51    59.6&      -4      38  10.2&   14.604&    2.091&  0.20&EB &\\
lra2a\_01462& 06524744-0359531  &    6      52    47.4&      -3      59  53.2&   12.295&    0.129&  0.04&DSCT &\\
lra2a\_01472& 06532070-0335031  &    6      53    20.7&      -3      35   2.9&   15.667&    4.634&  0.10&VAR &\\
lra2a\_01487& 06524919-0405332  &    6      52    49.2&      -4       5  33.4&   15.009&    1.724&  0.15&DCEP/EW &\\
lra2a\_01492*& 06530180-0356043  &    6      53     1.8&      -3      56   4.5&   13.896&    1.344&  0.05&SP/ELL &\\
lra2a\_01510*& 06530390-0359478  &    6      53     3.9&      -3      59  48.0&   14.123&   12.578&  0.05&$\alpha^2$CVn &\\
lra2a\_01533*&  	06525598-0411048 &    6      52    56.0&      -4      11   5.0&   15.110&    1.247&  0.06&$\gamma$Dor &\\
lra2a\_01550& 06534316-0335545  &    6      53    43.1&      -3      35  54.1&   12.602&    1.213&  0.08&VAR &\\
lra2a\_01551*& 	06525515-0415177  &    6      52    55.2&      -4      15  17.9&   14.796&    1.849&  0.10&VAR &\\
lra2a\_01565& 	06534650-0337257  &    6      53    46.4&      -3      37  25.3&   16.391&    3.229&  0.25&DCEP &\\
lra2a\_01573&  06535554-0333262 &    6      53    55.4&      -3      33  25.3&   15.641&    1.886&  0.20&DCEP &\\
lra2a\_01584*& 06530293-0418533  &    6      53     3.0&      -4      18  53.4&   15.835&    1.648&  0.10&$\alpha^2$CVn &\\
lra2a\_01592&06534626-0345272   &    6      53    46.2&      -3      45  27.2&   14.126&    2.276&  0.08&$\alpha^2$CVn &\\
lra2a\_01593*& 06525199-0429498  &    6      52    52.0&      -4      29  49.8&   16.380&    0.964&  0.60&EB &\\
lra2a\_01595&06534626-0345435   &    6      53    46.2&      -3      45  43.4&   15.574&    2.790&  0.35&DCEP &\\
lra2a\_01609*& 	06525078-0433067  &    6      52    50.8&      -4      33   6.7&   15.539&    0.227&  0.30&DSCT/EW &\\
lra2a\_01627*& 06525914-0429063  &    6      52    59.1&      -4      29   6.3&   16.564&    0.253&  0.56&EW &\\
lra2a\_01686& 	06540969-0344074  &    6      54     9.7&      -3      44   6.9&   15.452&    7.362&  0.10&LP &\\
lra2a\_01702&  06533647-0413215 &    6      53    36.8&      -4      13  15.6&   14.357&    0.876&  1.00& EA &XZ Mon\\
lra2a\_01739& 06530708-0440366  &    6      53     7.1&      -4      40  36.5&   15.650&    1.132&  0.20&EA &\\
lra2a\_01811&  	06535071-0412227 &    6      53    50.7&      -4      12  23.0&   17.030&    0.255&  0.65&EW &\\
lra2a\_01820c&  	06535998-0405514 &    6      53    60.0&      -4       5  51.7&   13.866&    0.498&  0.06&VAR &\\
lra2a\_01828c&06542227-0348393   &    6      54    22.2&      -3      48  38.7&   12.545&    0.498&  0.04&VAR &\\
lra2a\_01858c& 	06540106-0408135  &    6      54     1.1&      -4       8  13.8&   13.896&    0.498&  0.08&VAR &\\
\end{longtable}
}
\end{landscape}
}

\newpage
\renewcommand{\footnoterule}{}
\longtabL{5}{
\begin{landscape}
\small{
\begin{longtable}{lcllccccc}
\caption{\label{tabres2} Periodic variable stars detected in LRa2b.}\\ 
\hline\hline
BEST ID & 2MASS ID &$\alpha$(J2000) & $\delta$(J2000) & Mean mag(mag) & Period (days) & Amplitude (mag) & Type & Other Names\\
\hline
\endfirsthead
\caption{continued.}\\
\hline\hline
BEST ID  & 2MASS ID &$\alpha$(J2000) & $\delta$(J2000) & Mean mag(mag) & Period (days) & Amplitude (mag) & Type & Other Names\\
\hline
\endhead
\hline
\endfoot
lra2b\_00049& 06535497-0553063     & 6      53    55.0&      -5      53   6.6&   14.806&    6.269&  0.08&$\alpha^2$CVn &\\
lra2b\_00053& 06545012-0507166     & 6      54    50.1&      -5       7  16.8&   16.934&    0.280&  0.54&EW &\\
lra2b\_00116& 06531502-0617280    &  6      53    15.0&      -6      17  28.2&   13.296&    1.891&  0.12&EW &\\
lra2b\_00117& 06533942-0557263    &  6      53    39.4&      -5      57  26.7&   14.116&    0.886&  0.15&RRLyr &\\
lra2b\_00124& 06531766-0614104    &  6      53    17.7&      -6      14  10.6&   15.493&    8.936&  0.15&VAR &\\
lra2b\_00137& 06533463-0558279    &  6      53    34.7&      -5      58  28.2&   17.231&    0.708&  0.60&EB &\\
lra2b\_00179*\footnote{Stars with asterisk are located in the CoRoT's exoplanetary field-of-view.}& 06533609-0551510   &   6      53    36.1&      -5      51  51.2&   16.139&    2.568&  0.15&VAR &\\
lra2b\_00186& 	06530929-0612448    &  6      53     9.3&      -6      12  44.9&   12.294&    0.690&  0.10&EW &\\
lra2b\_00190& 	06543360-0503138    &  6      54    33.7&      -5       3  13.3&   13.171&    0.287&  0.09&DSCT &\\
lra2b\_00222&  06531071-0605587   &  6      53    10.7&      -6       5  58.9&   16.008&    4.290&  0.15&VAR &\\
lra2b\_00250& 06532919-0547016     & 6      53    29.2&      -5      47   1.9&   13.527&    1.022&  0.15&$\gamma$Dor &\\
lra2b\_00269& 	06535058-0524204     & 6      53    50.6&      -5      24  20.3&   16.148&    3.600&  0.40&EB &\\
lra2b\_00295&  	06541558-0458539    & 6      54    15.6&      -4      58  54.0&   15.369&    2.396&  0.20&DCEP &\\
lra2b\_00300&   	06540467-0506508   & 6      54     4.7&      -5       6  51.1&   16.351&    0.251&  0.40&EW &\\
lra2b\_00323*& 06531961-0540496    &  6      53    19.6&      -5      40  49.6&   15.383&    0.373&  0.70&EW &\\
lra2b\_00331& 06541082-0456551     & 6      54    10.8&      -4      56  55.2&   16.108&    0.344&  0.50&EB &\\
lra2b\_00348& 06524623-0601086     & 6      52    46.2&      -6       1   8.8&   15.318&    1.209&  0.15&DCEP &\\
lra2b\_00376*& 06530026-0542399    &  6      53     0.3&      -5      42  40.1&   12.489&    0.898&  0.05&VAR &\\
lra2b\_00378*& 	06530722-0536314    &  6      53     7.2&      -5      36  31.6&   15.789&   21.515&  0.20&VAR &\\
lra2b\_00389*& 06534165-0502400    &  6      53    41.7&      -5       2  40.1&   16.581&    2.254&  0.40&EA &\\
lra2b\_00396& 06524244-0549140     & 6      52    42.4&      -5      49  14.2&   13.283&    4.217&  0.08&ELL/SP &\\
lra2b\_00397&  06520502-0619199    & 6      52     5.0&      -6      19  19.9&   15.069&    1.101&  0.10&ELL/SP &\\
lra2b\_00402*& 	06524739-0544130    &  6      52    47.4&      -5      44  13.2&   13.247&    4.449&  0.10&ELL/SP &\\
lra2b\_00411*& 06533975-0458220    &  6      53    39.8&      -4      58  22.2&   14.255&    2.196&  0.07&EB &\\
lra2b\_00412*& 06530095-0529521    &  6      53     1.0&      -5      29  52.2&   13.485&    2.130&  0.05&ELL &\\
lra2b\_00419*& 06533531-0459416    &  6      53    35.3&      -4      59  41.7&   14.017&    1.368&  0.06&$\alpha^2$CVn &\\
lra2b\_00421*& 	06533499-0459421    &  6      53    35.0&      -4      59  42.2&   13.998&    1.369&  0.06&$\alpha^2$CVn &\\
lra2b\_00429&  06523176-0550078    & 6      52    31.8&      -5      50   8.0&   14.256&    2.117&  0.08&VAR &\\
lra2b\_00430*& 	06524238-0541250    &  6      52    42.4&      -5      41  25.1&   14.927&    0.352&  0.70&EW &\\
lra2b\_00443& 06520504-0608578     & 6      52     5.0&      -6       8  57.9&   15.792&    0.991&  0.60&EW &\\
lra2b\_00446& 06520464-0608504     & 6      52     4.7&      -6       8  50.9&   17.155&    0.991&  1.00&VAR &\\
lra2b\_00465& 06521691-0554377     & 6      52    16.9&      -5      54  37.7&   14.306&   14.355&  0.06&LP &\\
lra2b\_00466& 06520721-0602299     & 6      52     7.2&      -6       2  30.1&   15.829&    1.429&  0.40&EA &\\
lra2b\_00468& 	06522936-0544153     & 6      52    29.4&      -5      44  15.4&   14.252&    0.561&  0.20&EW &\\
lra2b\_00469& 06522734-0545516     & 6      52    27.3&      -5      45  51.8&   13.481&    0.911&  0.40& EA &EI Mon\\
lra2b\_00490&  	06520377-0601463    & 6      52     3.8&      -6       1  46.4&   14.639&    0.778&  0.10&RRLyr &\\
lra2b\_00493*& 	06524754-0525438    &  6      52    47.5&      -5      25  43.8&   13.142&    2.138&  0.08&$\alpha^2$CVn &\\
lra2b\_00506*&  	06522377-0542400   &  6      52    23.8&      -5      42  40.2&   14.371&    2.674&  0.04&$\alpha^2$CVn &\\
lra2b\_00511*& 06525612-0515414    &  6      52    56.1&      -5      15  41.6&   14.302&    0.620&  0.08&EW &\\
lra2b\_00513*& 	06524744-0522276    &  6      52    47.4&      -5      22  27.6&   14.758&    5.042&  0.05&VAR &\\
lra2b\_00514& 06515497-0605171     & 6      51    55.0&      -6       5  17.1&   14.513&    0.582&  0.38&EB &\\
lra2b\_00518&  06524748-0521590    & 6      52    47.5&      -5      21  59.1&   14.069&    0.528&  0.15&EW &\\
lra2b\_00527&  	06515947-0559171    & 6      51    59.5&      -5      59  17.4&   15.528&    2.718&  0.25&DCEP &\\
lra2b\_00551&  	06520404-0551036    & 6      52     4.0&      -5      51   3.6&   15.499&    0.354&  0.66&EW &\\
lra2b\_00566&  06520099-0551089    & 6      52     1.0&      -5      51   8.8&   16.870&    0.302&  0.64&EW &\\
lra2b\_00570*&  	06522020-0535137   &  6      52    20.2&      -5      35  13.7&   15.063&    1.260&  0.15&DCEP &\\
lra2b\_00590*&  06525605-0503355   &  6      52    56.1&      -5       3  35.6&   16.898&   24.786&  0.00&LP &\\
lra2b\_00595*& 	06530906-0452194    &  6      53     9.1&      -4      52  19.4&   14.352&    2.723&  0.06&SP &\\
lra2b\_00598*&  06523900-0515585   &  6      52    39.0&      -5      15  58.6&   14.903&    3.274&  0.08&$\alpha^2$CVn &\\
lra2b\_00617&  06512904-0609407    & 6      51    29.0&      -6       9  40.9&   16.923&    0.311&  0.53&EW &\\
lra2b\_00620&  06512696-0610504    & 6      51    27.0&      -6      10  50.3&   12.975&    1.683&  0.10&$\alpha^2$CVn &\\
lra2b\_00629*&  06521553-0529554   &  6      52    15.5&      -5      29  55.6&   14.415&    0.114&  0.08&DSCT &\\
lra2b\_00656& 	06512963-0602256     & 6      51    29.6&      -6       2  25.6&   14.434&    0.493&  0.08&SP &\\
lra2b\_00674*& 06522265-0516311    &  6      52    22.7&      -5      16  31.3&   13.419&    0.217&  0.05&SXPHE &\\
lra2b\_00675*& 06523346-0507400    &  6      52    33.5&      -5       7  39.8&   14.024&   32.080&  0.00&LP &\\
lra2b\_00678&  06511451-0610357    & 6      51    14.5&      -6      10  35.9&   12.724&   18.657&  0.07&$\alpha^2$CVn &\\
lra2b\_00681*&  06520776-0526248   &  6      52     7.8&      -5      26  24.8&   14.339&    2.047&  0.30&EA &\\
lra2b\_00687&  	06511964-0605018    & 6      51    19.7&      -6       5   2.1&   15.818&    2.274&  0.15&DCEP &\\
lra2b\_00722*& 	06523066-0458166    &  6      52    30.7&      -4      58  16.7&   17.249&    1.523&  0.30&DCEP &\\
lra2b\_00730*& 	06514351-0535028    &  6      51    43.5&      -5      35   3.0&   13.384&    2.674&  0.05&VAR &\\
lra2b\_00733*&     &  6      51    47.4&      -5      30  42.7&   15.918&    0.973&  0.14&RRLyr &\\
lra2b\_00738& 06512387-0549233     & 6      51    23.9&      -5      49  23.6&   13.585&    1.014&  0.12&SP &\\
lra2b\_00742*& 06513994-0535082    & 6      51    39.9&      -5      35   8.3&   14.805&    1.662&  0.06&SP &\\
lra2b\_00743& 06510486-0603305     & 6      51     4.9&      -6       3  30.5&   15.054&    1.140&  0.07&$\gamma$Dor &\\
lra2b\_00745& 	06512019-0550241     & 6      51    20.2&      -5      50  24.2&   16.239&    1.633&  0.18&DCEP &\\
lra2b\_00755*& 06520507-0511481    &  6      52     5.1&      -5      11  48.3&   14.990&    0.895&  0.14&RRLyr &\\
lra2b\_00761*& 	06515838-0515216    &  6      51    58.4&      -5      15  21.7&   15.134&   14.554&  0.30&DCEP &\\
lra2b\_00772*& 06514236-0525587    &  6      51    42.4&      -5      25  58.9&   13.694&    1.011&  0.03&$\gamma$Dor &\\
lra2b\_00785& 06505928-0557488     & 6      50    59.3&      -5      57  49.0&   13.949&    1.051&  0.06&VAR &\\
lra2b\_00786*& 	06521123-0458477    &  6      52    11.2&      -4      58  47.7&   15.476&   11.531&  0.15&ELL &\\
lra2b\_00800& 06505581-0558198     & 6      50    55.8&      -5      58  20.0&   14.887&    0.946&  0.15&EA &\\
lra2b\_00803&  06504824-0604034    & 6      50    48.2&      -6       4   3.4&   13.564&    0.300&  0.05&EW &\\
lra2b\_00807*&  	06515425-0509363   &  6      51    54.3&      -5       9  36.6&   15.881&    2.923&  0.16&DCEP &\\
lra2b\_00821*& 06513340-0524050    &  6      51    33.4&      -5      24   5.0&   15.484&    1.723&  0.15&EA &\\
lra2b\_00843& 06510251-0546453     & 6      51     2.5&      -5      46  45.5&   17.573&    0.737&  0.90&EB &\\
lra2b\_00844& 	06505996-0548080	     & 6      50    60.0&      -5      48   8.1&   16.407&    2.323&  0.17&SP &\\
lra2b\_00855*& 	06514089-0513051    &  6      51    40.9&      -5      13   5.3&   13.155&    2.208&  0.06&$\alpha^2$CVn &\\
lra2b\_00865&  	06504363-0557114    & 6      50    43.6&      -5      57  11.7&   14.758&   35.685&  0.00&LP &\\
lra2b\_00884*& 	06512729-0517154    &  6      51    27.3&      -5      17  15.6&   14.332&    0.897&  0.06&SP &\\
lra2b\_00915&  	06514570-0457314	    & 6      51    45.7&      -4      57  31.7&   15.885&    0.326&  0.40&EW &\\
lra2b\_00928&  06514146-0459244    & 6      51    41.5&      -4      59  24.8&   15.133&    0.656&  0.18&EA &\\
lra2b\_00949*& 06505007-0536311    &  6      50    50.1&      -5      36  31.3&   13.904&    1.862&  0.06&VAR &\\
lra2b\_00951*& 	06514300-0453055    &  6      51    43.0&      -4      53   5.7&   13.707&   10.812&  0.08&VAR &\\
lra2b\_00958*& 06505816-0526537    &  6      50    58.2&      -5      26  53.9&   16.071&    0.644&  0.30&EW &\\
lra2b\_00962*& 06513913-0452228    &  6      51    39.2&      -4      52  22.8&   14.773&    0.118&  0.06&DSCT &\\
lra2b\_00963*& 06505216-0530253    &  6      50    52.2&      -5      30  25.5&   16.299&    0.281&  0.30&EW &\\
lra2b\_00968*& 	06505364-0527405    &  6      50    53.6&      -5      27  40.8&   12.419&    0.975&  0.22&VAR &\\
lra2b\_00973*& 06504607-0532038    &  6      50    46.1&      -5      32   3.8&   16.684&    1.218&  0.20&EA &\\
lra2b\_00977*& 06504580-0531555    &  6      50    45.8&      -5      31  55.6&   15.448&    1.218&  0.19&EA &\\
lra2b\_00980*& 	06510243-0517248    &  6      51     2.4&      -5      17  25.1&   15.037&    3.724&  0.06&$\alpha^2$CVn &\\
lra2b\_00981&  	06501663-0554437    & 6      50    16.6&      -5      54  43.9&   14.572&    7.399&  0.06&VAR &\\
lra2b\_00988*& 06502912-0542443    &  6      50    29.1&      -5      42  44.5&   13.337&    4.746&  0.13&DCEP &\\
lra2b\_00992*& 	06504360-0530168    &  6      50    43.6&      -5      30  16.9&   14.879&    3.240&  0.12&$\alpha^2$CVn &\\
lra2b\_01014*& 	06512189-0455005   &   6      51    21.9&      -4      55   0.5&   16.412&    1.223&  0.21&EW &\\
lra2b\_01032*& 06504096-0525512	   &   6      50    41.0&      -5      25  51.3&   13.817&    1.266&  0.08&ELL &\\
lra2b\_01041& 06494574-0607317    &  6      49    45.7&      -6       7  31.5&   13.593&   13.050&  0.06&VAR &\\
lra2b\_01044c\footnote{Potential stellar crowding affecting the light curve of stars is marked with $c$.}& 	06501028-0547043    &  6      50    10.3&      -5      47   4.5&   14.918&    0.464&  0.07&RRLyr/SP &\\
lra2b\_01045c&  	06500991-0547067   &  6      50     9.9&      -5      47   6.9&   14.702&    0.464&  0.08&RRLyr/SP &\\
lra2b\_01046*&  06501652-0541335  &   6      50    16.5&      -5      41  33.7&   13.596&    2.385&  0.05&VAR &\\
lra2b\_01047&  	06500958-0547029   &  6      50     9.6&      -5      47   3.5&   15.212&    0.929&  0.14&VAR &\\
lra2b\_01052*& 06503814-0521479   &   6      50    38.1&      -5      21  48.0&   15.736&   21.404&  0.17&DCEP &\\
lra2b\_01074*& 06504774-0508275   &   6      50    47.8&      -5       8  27.7&   13.844&    0.634&  0.08&SP &\\
lra2b\_01075*& 06503217-0521030   &   6      50    32.2&      -5      21   2.9&   17.094&    1.551&  0.40&EW &\\
lra2b\_01076&  06494076-0602507   &  6      49    40.7&      -6       2  50.7&   16.192&    0.647&  0.31&EB &\\
lra2b\_01080*&    &   6      50    23.9&      -5      26  29.5&   15.488&    0.746&  0.18&EA &\\
lra2b\_01081*& 06510815-0449172   &   6      51     8.2&      -4      49  17.3&   13.373&    8.074&  0.10&$\alpha^2$CVn &\\
lra2b\_01082&   	06495929-0545296  &  6      49    59.3&      -5      45  29.7&   16.376&    0.671&  0.60&EA &\\
lra2b\_01087&   	06495406-0549242  &  6      49    54.1&      -5      49  24.3&   16.489&    0.532&  0.40&EW &\\
lra2b\_01088&   06495901-0545215  &  6      49    59.0&      -5      45  21.7&   16.034&    0.671&  0.10&EA &\\
lra2b\_01098&  06493916-0600049   &  6      49    39.1&      -6       0   5.1&   14.808&    9.771&  0.20&DCEP/CV &\\
lra2b\_01100&  06495722-0544422   &  6      49    57.2&      -5      44  42.3&   15.000&    7.347&  0.19&DCEP &\\
lra2b\_01113*& 06501060-0531341   &   6      50    10.6&      -5      31  34.3&   13.444&    0.719&  0.05&RRLyr &\\
lra2b\_01126*c&  06501813-0523164  &   6      50    18.1&      -5      23  16.4&   16.336&    1.287&  0.15&DCEP &\\
lra2b\_01127*c&  06501791-0523109  &   6      50    18.0&      -5      23  10.5&   16.474&    1.287&  0.15&DCEP &\\
lra2b\_01139&   06494359-0549498  &  6      49    43.6&      -5      49  49.8&   13.666&    2.656&  0.08&SP &\\
lra2b\_01144*&  	06504492-0459068  &   6      50    44.9&      -4      59   7.1&   13.625&    0.586&  0.20&EB &\\
lra2b\_01149*&  06500103-0534332  &   6      50     1.0&      -5      34  33.3&   15.151&    0.906&  0.10&EW &\\
lra2b\_01151*& 06503379-0507263   &   6      50    33.8&      -5       7  26.4&   14.743&    2.184&  0.12&DCEP &\\
lra2b\_01172*& 06500956-0524553   &   6      50     9.6&      -5      24  55.4&   13.846&    0.618&  0.08&RRLyr &\\
lra2b\_01192*& 06504495-0453327   &   6      50    45.0&      -4      53  32.7&   14.736&    3.878&  0.10&EA &\\
lra2b\_01202*&  	06500738-0521515  &   6      50     7.4&      -5      21  51.6&   15.698&   11.039&  0.14&DCEP &\\
lra2b\_01205&  	06494101-0542384   &  6      49    41.0&      -5      42  38.7&   13.920&    0.413&  0.15&EB &\\
lra2b\_01207*&  06500183-0524501  &   6      50     1.8&      -5      24  50.2&   14.341&    1.958&  0.05&SP &\\
lra2b\_01208*& 06500867-0518563   &   6      50     8.7&      -5      18  56.4&   15.807&    9.226&  0.20&VAR &\\
lra2b\_01221*& 06495556-0527141   &   6      49    55.6&      -5      27  14.2&   15.767&    0.832&  0.16&EA &\\
lra2b\_01225*& 06503112-0457068   &   6      50    31.1&      -4      57   7.0&   14.154&    1.659&  0.08&VAR &\\
lra2b\_01227&   06492270-0552235  &  6      49    22.7&      -5      52  23.8&   16.377&   27.914&  0.00&LP &\\
lra2b\_01248*& 06495338-0522213   &   6      49    53.4&      -5      22  21.4&   16.321&    2.889&  0.25&DCEP &\\
lra2b\_01256*&  	06493107-0538073  &   6      49    31.1&      -5      38   7.3&   16.366&    3.036&  0.23&DCEP &\\
lra2b\_01257&  06491567-0550387   &  6      49    15.7&      -5      50  38.9&   14.070&    1.486&  0.08&EA &\\
lra2b\_01264&   	06491528-0548185  &  6      49    15.3&      -5      48  18.6&   16.182&    0.265&  0.30&DSCT &\\
lra2b\_01265&  06502464-0450471   &  6      50    24.7&      -4      50  47.3&   15.058&    1.294&  0.06&SP &\\
lra2b\_01274*&  	06493709-0525568  &   6      49    37.1&      -5      25  56.5&   14.714&    6.168&  0.08&$\alpha^2$CVn &\\
lra2b\_01275c&   06491749-0541441  &  6      49    17.5&      -5      41  44.2&   15.400&    0.260&  0.30&EW &\\
lra2b\_01277c&   06491712-0541423  &  6      49    17.1&      -5      41  42.6&   15.395&    0.260&  0.30&EW &\\
lra2b\_01285*&  06500301-0502506  &   6      50     3.0&      -5       2  50.7&   16.109&    1.541&  0.33&EB &\\
lra2b\_01286*& 06493048-0529167   &   6      49    30.5&      -5      29  16.7&   15.190&    3.194&  0.17&DCEP &\\
lra2b\_01287c&  	06490818-0546524   &  6      49     8.2&      -5      46  52.8&   15.826&    1.021&  0.10&$\gamma$Dor &\\
lra2b\_01288c&     &  6      49     8.3&      -5      46  48.5&   16.031&    1.021&  0.15&$\gamma$Dor &\\
lra2b\_01292*c&  06501084-0455094  &   6      50    10.8&      -4      55   9.4&   14.300&    2.959&  0.12&SP &\\
lra2b\_01293*c&  06501055-0455148  &   6      50    10.6&      -4      55  14.2&   14.527&    2.956&  0.12&SP &\\
lra2b\_01296&     &  6      49    30.9&      -5      27  16.1&   15.472&    7.616&  0.08&VAR &\\
lra2b\_01303*&   06503162-0436431 &   6      50    31.6&      -4      36  42.7&   16.194&    2.135&  0.18&DCEP &\\
lra2b\_01307*&  06494075-0518003  &   6      49    40.7&      -5      18   0.3&   15.074&    4.091&  0.10&DCEP &\\
lra2b\_01309*&  06494036-0518019  &   6      49    40.4&      -5      18   2.0&   15.138&    4.094&  0.10&DCEP &\\
lra2b\_01316*& 06494150-0515029   &   6      49    41.5&      -5      15   3.0&   15.147&    4.197&  0.06&VAR &\\
lra2b\_01330&   06490566-0540459  &  6      49     5.7&      -5      40  46.0&   14.815&    1.412&  0.08&EB &\\
lra2b\_01334*&  	06495851-0456485  &   6      49    58.5&      -4      56  48.7&   16.261&    0.303&  0.60&EW &\\
lra2b\_01340*&  06492045-0526540  &   6      49    20.5&      -5      26  54.1&   14.049&    1.266&  0.30&EB &\\
lra2b\_01352*&  06492382-0521252  &   6      49    23.8&      -5      21  25.3&   15.384&    1.584&  0.12&VAR &\\
lra2b\_01364*&  06493845-0508114  &   6      49    38.4&      -5       8  11.4&   14.225&    4.751&  0.15&DCEP &\\
lra2b\_01368*&  06500269-0447453  &   6      50     2.7&      -4      47  45.5&   14.586&    1.190&  0.10&SP &\\
lra2b\_01369&   	06484618-0549328  &  6      48    46.2&      -5      49  32.9&   12.352&    0.117&  0.06&SXPHE &\\
lra2b\_01371*& 06493408-0510099   &   6      49    34.1&      -5      10   9.9&   14.500&    2.853&  0.08&EA &\\
lra2b\_01372*&    &   6      49    34.3&      -5      10   4.4&   15.016&    2.856&  0.12&EA &\\
lra2b\_01384&  06484661-0545492   &  6      48    46.6&      -5      45  49.5&   14.932&    0.470&  0.40&EW &\\
lra2b\_01387&     &  6      48    46.2&      -5      45  45.9&   15.928&    0.470&  0.80& VAR &\\
lra2b\_01394&  06483196-0556002   &  6      48    31.9&      -5      56   0.6&   15.239&    4.075&  0.10&VAR &\\
lra2b\_01396&  06484147-0547398   &  6      48    41.5&      -5      47  40.0&   17.123&    1.526&  0.48&DCEP &\\
lra2b\_01404&  06482945-0556226   &  6      48    29.4&      -5      56  22.9&   15.166&    0.943&  0.15&EB &\\
lra2b\_01405&  06483132-0554408   &  6      48    31.3&      -5      54  41.0&   13.455&   20.097&  0.04&LP &\\
lra2b\_01437&  06484069-0537534   &  6      48    40.7&      -5      37  53.5&   14.378&    0.418&  0.40&EW &\\
lra2b\_01458*c&    &   6      49    15.5&      -5       3  23.4&   14.816&    0.291&  0.60&EW &\\
lra2b\_01459*c& 06491524-0503301   &   6      49    15.2&      -5       3  30.4&   13.860&    0.291&  0.60&EW &\\
lra2b\_01463*& 	06485685-0518103   &   6      48    56.9&      -5      18  10.4&   14.986&    1.110&  0.14&EA &\\
lra2b\_01482*& 	06482851-0535595	   &   6      48    28.5&      -5      35  59.7&   13.055&    1.918&  0.06&ELL &\\
lra2b\_01485& 	06480100-0557326    &  6      48     1.0&      -5      57  33.0&   12.295&    0.497&  0.40&EB &\\
lra2b\_01511*& 06485620-0507196   &   6      48    56.2&      -5       7  19.7&   14.946&    1.020&  0.08&SP &\\
lra2b\_01549& 06475182-0552172    &  6      47    51.8&      -5      52  17.7&   14.762&    1.588&  0.07&ELL &\\
lra2b\_01552*& 	06483819-0513562   &   6      48    38.2&      -5      13  56.1&   16.363&    0.280&  0.40&EB &\\
lra2b\_01554*& 06481923-0528506   &   6      48    19.2&      -5      28  50.6&   14.868&   18.604&  0.10&VAR &\\
lra2b\_01561*& 06490490-0449310   &   6      49     4.9&      -4      49  31.2&   16.145&    0.363&  0.60&EW &\\
lra2b\_01600*& 	06484535-0458261   &   6      48    45.3&      -4      58  26.2&   15.709&    0.387&  0.30&EA &\\
lra2b\_01630*& 06484143-0456395   &   6      48    41.4&      -4      56  39.6&   15.547&    5.893&  0.60&EA &\\
lra2b\_01648&  	06474385-0541125   &  6      47    43.8&      -5      41  12.8&   12.560&    1.952&  0.12&ELL &\\
lra2b\_01660& 	06472622-0554256    &  6      47    26.2&      -5      54  26.1&   13.948&    0.894&  0.06&SP &\\
lra2b\_01691&  	06474781-0534160   &  6      47    47.8&      -5      34  16.2&   16.389&    2.230&  0.20&DCEP &\\
\end{longtable}
}
\end{landscape}
}


\begin{thebibliography}{}
  
\bibitem[Ahnert 1949]{Ahnert1949}Ahnert, P., 1949, VSS 1, N3

\bibitem[Alard \& Lupton 1998]{Alard1998} Alard, C. \& Lupton, R. H. $1998$, \apj, 503, 325

\bibitem[Baglin et al. 1998]{baglin} Baglin, A., Auvergne, M., Barge, P., Buey, J.-T., Catala, C., Michel, E., Weiss, W., \& CoRoT Team 2002, ESA SP-485: Stellar Structure and Habitable Planet Finding, 17 


\bibitem[Deleuil et al. 2006]{Deleuil2006} Deleuil, M., Moutou, C., Deeg, H., J., Meunier, J., C., Surace, C., Guterman, P., Almenara, J. M., Alonso, R., Barge, P., Bouchy, F., Erikson, A., Leger, A., Loeillet, B., Ollivier, M., Pont, F., Rauer, H., Rouan, D., and Queloz, D. $2006$, ESA SP-1306: The CoRoT Mission, 341


\bibitem[Eyer \& Mowlavi (2008)]{Eyer} Eyer, L. \& Mowlavi, N. 2008, Journal of Physics: Conference Series 118 

\bibitem[Fernandez-Figueroa et al. 1994]{Fernandez1994}Fernandez-Figueroa, M., Montes, D., De Castro, E., \& Cornide, M. 1994, \apjs, 90, 433


\bibitem[Kaye et al. 1999]{kaye1999}Kaye, B., A., Handler, G., Krisciunas, K. \& Poretti, E., Zerbi, M., F. 1999, \pasp, 111, 840  

\bibitem[Kabath et al. (2007)]{Kabath2007} Kabath, P., Eigm\"{u}ller, P., Erikson, A., Hedelt, P., Rauer, H., Titz, R., Wiese, T., Karoff, C. $2007$, \aj, 134, 1560

\bibitem[Kabath et al. 2008]{Kabath2008} Kabath, P., Eigm\"{u}ller, P., Erikson, A., Hedelt, P., von Paris, P.,  Rauer, H., Renner, S., Titz, R., Karoff, C. $2008$, \aj, 136, 654

\bibitem[Kabath et al. 2009]{Kabath2009} Kabath, P., Fruth, T., Rauer, H., Erikson, A., Murphy, M. G., Chini, R., Lemke, R., Csizmadia, Sz., Eigm\"{u}ller, P., Pasternacki, T., Titz, R $2009$, \aj, 137, 3911

\bibitem[Karoff et al. (2007)]{Karoff2007} Karoff, C., Rauer, H., Erikson, A., Voss, H., Deleuil, M., Moutou, C., Meunier, J., C., Deeg, H.  $2007$, \aj, 134, 766

\bibitem[Maceroni \& van't Veer 1993]{Oconnel}Maceroni, C., van't Veer, F. 1993, \aap, 277, 515


\bibitem[P\'{a}l \& Bakos 2007]{Pal2007} P\'{a}l, A., Bakos, G. \'{A}., $2007$, ASPC, 366, 340

\bibitem[Peraiah 1998]{Pariah1998} Peraiah, A., Srinivasa Rao, M. $1998$, \aaps, 132, 45   

\bibitem[Pojmanski 2002]{Pojmanski2002} Pojmanski, G., $2002$, AcA, 52, 397
                    
\bibitem[Poretti et al. (2005)]{poretti}Poretti, E., Alonso, R., Amado, P. J., Belmonte, J. A., Garrido, R., Mart\'{i}n-Ruiz, S., Uytterhoeven, K., Catala, C., Lebreton, Y., Michel, E., Su\'{a}rez, J. C., Aerts, C., Creevey, O., Goupil, M. J., Mantegazza, L., Mathias, P., Rainer, M., Weiss, W. W. $2005$, \aj, 129, 246

\bibitem[Rauer et al. 2004]{Rauer2004} Rauer, H., Eisl\"{o}ffel, J., Erikson, A., Guenther, E., Hatzes, A. P., Michaelis, H., Voss, H. $2004$, \pasp, $116$, $38$

\bibitem[Rauer et al. 2008]{Rauer2008a} Rauer, H., Fruth, T., Erikson, A. $2008a$, \pasp, 120, 852

\bibitem[Rauer et al. 2009]{Rauer2009} Rauer, H., Erikson, A., Kabath, P., Hedelt, P., Csizmadia, Sz., Eigm\"{u}ller, P., von Paris, P., Renner, S., Titz, R., Voss, H. $2009$, submitted to \aj

\bibitem[Schwarzenberg-Czerny 1996]{schwarzenberg-czerny} Schwarzenberg-Czerny, A. $1996$, \apj, 460, 107

\bibitem[(Skrutskie et al. 2006)]{Skrutskie2006}Skrutskie, M. F., Cutri, R. M., Stiening, R., Weinberg, M. D., Schneider, S., Carpenter, J. M., Beichman, C., Capps, R., Chester, T., Elias, J., Huchra, J., Liebert, J., Lonsdale, C., Monet, D. G., Price, S., Seitzer, P., Jarrett, T., Kirkpatrick, J. D., Gizis, J., Howard, E., Evans, T., Fowler, J., Fullmer, L., Hurt, R., Light, R., Kopan, E. L., Marsh, K. A., McCallon, H. L., Tam, R., Van Dyk, S., Wheelock, S. $2006$, \aj, 131, 1163

\bibitem[Stetson (1996)]{stetson} Stetson, P. B. $1996$, \pasp, 108, 851  
  
\bibitem[Sterken \& Jaschek 1996]{sterken} Sterken, C., Jaschek, C., $1996$, Light curves of variable stars, Cambridge University press

\bibitem[Tamuz et al. 2005]{Tamuz2005} Tamuz, O., Mazeh, T. and Zucker, S. $2005$, \mnras, 356, 1466

\bibitem[Udalski et al. 1997]{Udalski1997}Udalski, A., Kubiak, M., Szyma\'{n}ski, M. 1997, AcA, 47, 319

\bibitem[Vaz 1985]{Vaz1985}Vaz, L. P. R $1985$, \apss, 113, 2, 349

\bibitem[Zhang et al. (2003)]{zhang} Zhang, X., Deng, L., Xin, Y., \& Zhou, X. $2003$, Chinese Journal of Astronomy and Astrophysics, 3, 151
 
\end{thebibliography}
\end{document}